\documentclass[12pt,a4paper]{article}

\usepackage{mathtools} 
\usepackage{amssymb}
\usepackage{bm}
\usepackage{accents}
\usepackage{color}
\usepackage[bookmarks=true,hyperfigures=true]{hyperref}
\usepackage{graphicx}
\usepackage[nosort]{cite}
\usepackage[bulletsep]{collref}
\usepackage{tensor}
\usepackage[british]{babel}

\usepackage[a4paper,text={173mm,216mm},centering]{geometry}

\allowdisplaybreaks[3]

\newcommand{\nn}{\nonumber}

\usepackage[font=small,labelfont=bf,width=0.85\textwidth]{caption}

\let\oldbfseries=\bfseries
\renewcommand{\bfseries}{\oldbfseries\boldmath}

\newcommand{\eqn}[1]{(\ref{#1})}

\newcommand{\sfrac}[2]{{\textstyle\frac{#1}{#2}}}
\newcommand{\half}{\sfrac{1}{2}}
\newcommand{\ihalf}{\sfrac{i}{2}}

\newcommand{\quarter}{\sfrac{1}{4}}

\DeclareMathOperator{\tr}{tr}

\newcommand*{\diff}{{\mathrm d}}

\newcommand{\str}{\mathop{\mathrm{str}}}
\newcommand{\ft}[2]{{\textstyle\frac{#1}{#2}}}

\renewcommand{\a}{\alpha}
\renewcommand{\b}{\beta}
\newcommand{\g}{\gamma}

\newcommand{\da}{{\dot{\alpha}}}
\newcommand{\db}{{\dot{\beta}}}
\newcommand{\dg}{{\dot{\gamma}}}
\newcommand{\la}{\lambda}
\newcommand{\bla}{\bar{\lambda}}
\newcommand{\dbla}{\dot{\bar{\lambda}}}

\newcommand{\eps}{\epsilon}
\newcommand{\dx}{{\dot x}}
\newcommand{\s}{\sigma}
\newcommand{\bs}{\bar{\sigma}}
\newcommand{\btheta}{\bar\theta}
\newcommand{\vart}{\vartheta}
\newcommand{\bvart}{\bar\vartheta}

\newcommand{\Qb}{{\bar{Q}}}
\newcommand{\Sb}{{\bar{S}}}

\begin{document}
\thispagestyle{empty}
%\ifarxiv\vspace*{-20mm}\fi

\begingroup\raggedleft\footnotesize\ttfamily
HU-EP-15/31\
\vspace{15mm}
\endgroup

\begin{center}
{\Large\bfseries Bonus Symmetry for Super Wilson Loops\par}
\vspace{25mm}

\begingroup\scshape\large 
Hagen M\"unkler
\endgroup
\vspace{5mm}

\textit{Institut f\"ur Physik and IRIS Adlershof, Humboldt-Universit\"at zu Berlin, \phantom{$^\S$}\\
Zum Gro{\ss}en Windkanal 6, D-12489 Berlin, Germany} \\[0.1cm]
\texttt{ \small{ muenkler@physik.hu-berlin.de\phantom{\ldots}}} \\ \vspace{5mm}

\vspace{18mm}

\textbf{Abstract}\vspace{5mm}\par
\begin{minipage}{14.7cm}
The Yangian level-one hypercharge generator for the super Wilson loop in $\mathcal{N}=4$ supersymmetric Yang-Mills theory is constructed. Moreover, evidence for the presence of a corresponding symmetry generator at all higher levels is provided. The derivation is restricted to the strong-coupling description of the super Wilson loop and based on the construction of novel conserved charges for type IIB superstrings on $\mathrm{AdS}_5 \times \mathrm{S}^5$.
\end{minipage}\par
\end{center}
\newpage

\setcounter{tocdepth}{2}
\hrule height 0.75pt
\setcounter{tocdepth}{3}
\tableofcontents
\vspace{0.8cm}
\hrule height 0.75pt
\vspace{1cm}

\setcounter{tocdepth}{2}

\hyphenation{geo-me-trische}

%%%%%%%%%%%%%%%%%%%%%%%%%%%%%%%%%%%%%%%%%%%%%%%%%%%%%%%%%%%%%%%%%%%%%%%%%%%%%%%%
%%%%%%%%%%%%%%%%%%%%%%%%%%%%%%%%%%%%%%%%%%%%%%%%%%%%%%%%%%%%%%%%%%%%%%%%%%%%%%%%

\section{Introduction}

The last decade has witnessed spectacular progress towards exact solutions in planar, maximally supersymmetric Yang-Mills (SYM) theory. Much of this progress has been sparked by the use of integrability, which often comes with an infinite-dimensional symmetry of the Yangian type. To make further progress in this direction it is important to fully characterize and understand the symmetries that appear in planar $\mathcal{N}=4$ SYM theory. For the S-matrix, it has been argued that apart from the Yangian $Y[\mathfrak{psu}(2,2\vert4)]$ symmetries \cite{Drummond:2009fd,Beisert:2010gn,Sever:2009aa}, there is a so-called secret or bonus symmetry \cite{Matsumoto:2007rh,Beisert:2011pn}, the level-1 recurrence of the hypercharge generator, which is itself not a symmetry. Corresponding conserved charges have been constructed for the pure spinor superstring \cite{Berkovits:2011kn} in all odd levels of the Yangian. 

In this letter, we construct the level-one bonus symmetry for the super Wilson loop in $\mathcal{N}=4$ SYM theory at strong coupling and provide strong evidence for the existence of a corresponding symmetry generator at any higher level of the Yangian. The super Wilson loop is a generalization of the Maldacena-Wilson loop \cite{Maldacena:1998im}, which also includes the fermionic fields of $\mathcal{N}=4$ SYM. It has already been considered in the early days of the AdS/CFT correspondence \cite{Ooguri:2000ps} and recently a complete construction was given \cite{Beisert:2015jxa}. Hints for a Yangian symmetry of the super Wilson loop had been obtained before \cite{Muller:2013rta} and this symmetry has by now been established at weak \cite{Beisert:2015xxx} and strong coupling \cite{Munkler:2015gja}. The strong coupling description of the super Wilson loop is given by the minimal surface of a type IIB superstring ending on the conformal boundary superspace and the Yangian symmetry at strong coupling can be derived from explicitly computing the conserved charges that follow from the classical integrability of the string model. The conserved charges that lead to the bonus symmetry can be constructed in a central extension of the coset superspace of the type IIB superstring, as it was also considered in \cite{Berkovits:2011kn}. In the case at hand, the disk topology of the minimal surface allows for a simplified construction, which includes charges that lead to symmetries at any higher level of the Yangian.

Let us briefly explain how this paper is structured. The construction of the higher conserved charges is explained in section \ref{sec:charges} and the corresponding symmetry generator at level one is derived in section \ref{level1B}. The derivation of the symmetry generators at level one from the respective conserved charges has been discussed in detail in \cite{Munkler:2015gja} and the calculation performed there can easily be lifted to the centrally extended coset. The exposition in section \ref{level1B} is focused on the new symmetry generators that follow from the central extension and the reader is referred to \cite{Munkler:2015gja} for a detailed calculation as well as our conventions. The additional conventions and symmetry generators for the central extension are collected in the appendices \ref{app:u224} and \ref{app:generators}. We present concluding remarks in section \ref{sec:outlook} and give an outlook on possible future works.

\section{Conserved Charges of the \texorpdfstring{$\mathrm{AdS}_5\times \mathrm{S}^5$}{AdS5 x S5} - Superstring}
\label{sec:charges}
In this section, we construct non-local conserved charges for the type IIB superstring on $\mathrm{AdS}_5 \times \mathrm{S}^5$. We consider the case of a minimal surface, where the world-sheet is Euclidean and has the topology of a disk. The results can be transferred to a Minkowskian world-sheet, whereas the restriction to a disk topology is crucial. Our exposition  follows \cite{Arutyunov:2009ga}.

The type IIB superstring theory in $\mathrm{AdS_5} \times \mathrm{S}^5$ can be described by a sigma model type action with target space
\begin{align}
\frac{\mathrm{PSU}(2,2\vert 4)}{\mathrm{SO}(4,1) \times \mathrm{SO}(5)} \,. \label{targetspace}
\end{align}
It is often convenient to consider the coset space $\mathrm{SU}(2,2\vert4)/\left( \mathrm{SO}(4,1) \times \mathrm{SO}(5) \right)$, since $\mathrm{SU}(2,2\vert4)$ allows for a matrix representation. Concretely, the Lie superalgebra $\mathfrak{su}(2,2\vert4)$ is given by the set of $\left( 4 \vert 4 \right) \times \left( 4 \vert 4 \right)$ supermatrices which satisfy a hermiticity condition and have vanishing supertrace, see appendix~\ref{app:u224} for details. The Lie superalgebra $\mathfrak{psu}(2,2\vert4)$ is obtained from $\mathfrak{su}(2,2\vert4)$ by projecting out the central charge $C$, which commutes with all other generators. 

For a function $g(\tau, s) \in \mathrm{SU}(2,2\vert4)$ of the world-sheet coordinates, the Cartan form $A_i = - g^{-1} \partial_i g$ provides a flat connection, 
\begin{align}
\eps^{i j} \left( 2 \partial_i \, A_j - \left[A_i \, , \, A_j \right] \right) = 0 \, ,
\end{align}
taking values in $\mathfrak{su}(2,2\vert4)$. This algebra may be endowed with a $\mathbb{Z}_4$-grading: 
\begin{align}
\mathfrak{su}(2,2\vert4) &= \mathfrak{g}^{(0)} \oplus \mathfrak{g}^{(2)} \oplus \mathfrak{g}^{(1)} \oplus \mathfrak{g}^{(3)} \, , \qquad \qquad  \left[ \mathfrak{g}^{(k)} \, , \, \mathfrak{g}^{(l)} \right] \subset \mathfrak{g}^{(k+l) \, \mathrm{mod} 4 }  \,.
\end{align}
Here, $\mathfrak{g}^{(0)} \oplus \mathfrak{g}^{(2)}$ is the bosonic subalgebra of $\mathfrak{su}(2,2\vert4)$ and $\mathfrak{g}^{(1)}\oplus \mathfrak{g}^{(3)}$ comprises the fermionic generators. Based on the projection operators $P^{(k)}: \mathfrak{su}(2,2\vert4) \to \mathfrak{g}^{(k)}$ onto these graded components, we introduce the short-hand notation 
\begin{align*}
B^{(k)} = P^{(k)} \left( B \right) \, , \qquad B^{(1)\pm (3)} = B^{(1)} \pm B^{(3)} \, .
\end{align*}
More details on the $\mathbb{Z}_4$ decomposition of $\mathfrak{su}(2,2\vert4)$ can be found in appendix~\ref{app:u224}. In the fundamental representation of $\mathfrak{su}(2,2\vert4)$ a metric can be defined in terms of the supertrace, $G_{ab} = \str \left( T_a T_b \right)$. The superstring action can then be written as
\begin{align}
S = - \frac{\sqrt{\la}}{4 \pi}  \int \diff \tau \, \diff \s \left \lbrace \gamma ^{i j} \,  \str \left( A_i ^{(2)}  A_j ^{(2)} \right) + i \, \eps^{ij} \str \left( A_i ^ {(1)}  A_j ^ {(3)} \right) \right \rbrace \,. \label{action}
\end{align}
Here, $\gamma^{i j} = \sqrt{\det(h_{ij})} \, h^{ij}$ is manifestly Weyl-invariant and we fix the convention $\eps^{\tau s} = 1$. Note also, that we work with a Euclidean world-sheet metric -- resulting in the factor of $i$ in front of the fermionic term -- since the induced metric on the world-sheet is Euclidean for the boundary conditions we consider. Varying the action with respect to $g$ leads to
\begin{align*}
\delta S = - \frac{\sqrt{\la}}{4 \pi}  \int \diff \tau \, \diff \s \str \left( g^{-1} \delta g \left( \partial_i \Lambda ^i - \left[ A_i , \Lambda^i \right]  \right) \right) \, , \qquad \Lambda ^i &=  \gamma ^{i j} \, A_j ^ {(2)} - \sfrac{i}{2}  \, \eps^{i j} A_j^{(1)-(3)} \, ,
\end{align*}
and since $C$ is a singular vector in $\mathfrak{su}(2,2 \vert 4)$, we find the equations of motion 
\begin{align}
\partial_i \, \Lambda ^i - \left[ A_i \, , \, \Lambda ^i \right] = \alpha \, C \, .
\end{align}
Here, $\alpha$ is an arbitrary function of the world-sheet coordinates $\tau$ and $\s$. This shows that the degree of freedom associated with $C$ is spurious and the target space is really \eqn{targetspace}. However, for the study of conserved charges it will be interesting to keep $C$ in place. The variation with respect to the world-sheet metric gives the Virasoro constraints
\begin{align}
\str \left( A_i ^{(2)}  A_j ^{(2)} \right) - \half \gamma _{i j} \, \gamma ^{k l}  \str \left( A_k ^{(2)}  A_l ^{(2)} \right) = 0 \, .
\end{align}

We now turn to the construction of conserved charges, for which we apply the classical integrability of the string model\cite{Bena:2003wd}. As we are working over $\mathfrak{su}(2,2\vert4)$, the flatness condition for the Lax connection reads
\begin{align}
\eps^{ij} \left(2 \partial_i L_j [z] - \left[L_i [z] \, , \, L_j [z] \right] \right)= \tilde{\alpha}[z] \, C \, . 
\end{align}
The Lax connection and $\tilde{\alpha}$ are given by 
%where both $L_i$ and $\tilde{\alpha}$ depend on a spectral parameter $z$,
\begin{align}
%L_i &= A_i ^{(0)} + \frac{1+z^2}{1-z^2} \, A_i^{(2)} - \frac{2i \, z}{1-z^2} \, \gamma_{ij} \, \eps^{jk} \, A_k ^{(2)} + \frac{\sqrt{1+z}}{\sqrt{1-z}} \, A_i ^{(1)} + \frac{\sqrt{1-z}}{\sqrt{1+z}} \, A_i ^{(3)} \,. \\
L_i[z] &= A_i ^{(0)} + \frac{1+z^2}{1-z^2} \, A_i^{(2)} - \frac{2i \, z}{1-z^2} \, \gamma_{ij} \, \eps^{jk} \, A_k ^{(2)} + \frac{1}{\sqrt{1-z^2}} \, A_i ^{(1)+(3)} + \frac{z }{\sqrt{1-z^2}} \, A_i ^{(1)-(3)} \,. \\
\tilde{\alpha} [z] &= \frac{4i \, z}{1-z^2} \, \alpha (\tau , \s) \, .
\end{align}
The $z$-dependence of $\tilde{\alpha}$ follows from the $P^{(2)}$-projection of the flatness condition, which can be reduced to the respective projection of the equations of motion by using the flatness condition for $A_i$. Note in particular that the $z$-dependence of $\tilde{\alpha}$ is fixed completely, whereas its dependence on the world-sheet coordinates is arbitrary. In order to construct the conserved charges, we consider the gauge transformed Lax connection 
\begin{align}
l_i &= g L_i g^{-1} + \left(\partial_i g \right) g^{-1} = g \left( L_i - A_i \right) g^{-1} \notag \\
&= \frac{2z^2}{1-z^2} \, a_i^{(2)} - \frac{2i \, z}{1-z^2} \, \gamma_{ij} \, \eps^{jk} \, a_k ^{(2)} + \frac{1 - \sqrt{1-z^2}}{\sqrt{1-z^2}} \, a_i ^{(1)+(3)} + \frac{z }{\sqrt{1-z^2}} \, a_i ^{(1)-(3)} \, ,
\end{align}
where we defined $a_i ^{(k)} = g \, A_i ^{(k)} \, g^{-1}$. Note in particular that the $z$-expansion of $l_i= \sum l_{i , n}  z^n$ starts at order $z$. A tower of multi-local conserved charges \cite{Beisert:2005bm} can be extracted from expanding the monodromy matrix associated to the gauge transformed Lax connection around $z=0$. We use conformal gauge and consider a closed curve at constant value of $\tau$. The monodromy matrix is given by
\begin{align*}
T[z]= \mathcal{P}\exp{\left(\oint \diff \s\,  l_\s \right)} = \sum \limits _{n=0} ^\infty T_n \, z^n \, .
\end{align*} 
The evolution of the monodromy matrix with respect to the parameter $\tau$ is governed by
\begin{align}
\partial_\tau T[z] &= \int  _0 ^L \diff \s \, \left [  \mathcal{P} \exp{\int _\s ^L  \diff \s\,  l_\s } \right] \partial_\tau  l_\s \, \left [  \mathcal{P} \exp{\int _0 ^\s  \diff \s\,  l_\s } \right] \nn \\
&=  \int  _0 ^L \diff \s \, \partial_\s \left[  \left(  \mathcal{P} \exp{\int _\s ^L  \diff \s\,  l_\s } \right) l_\tau \, \left(  \mathcal{P} \exp{\int _0 ^\s  \diff \s\,  l_\s } \right) \right] +  \int  _0 ^L \diff \s \tilde{\alpha} C \left [  \mathcal{P} \exp{\int _0 ^L  \diff \s\,  l_\s } \right] \nn \\
% & \quad +  \int  _0 ^L \diff \s \tilde{\alpha} C \left [  \mathcal{P} \exp{\int _\s ^L  \diff \s\,  l_\s } \right] \left [  \mathcal{P} \exp{\int _0 ^\s  \diff \s\,  l_\s } \right] \nn \\ 
&= \big[ l_\tau (0, \tau , z) , T[z] \big] + \frac{4i z}{1-z^2}  T[z]  \int \diff \s \alpha C  \, . \label{monocons}
\end{align}
The expansion of $l_\tau$ starts at order $z$ and so we find $\partial_\tau T_1 = 4i \int \diff \s \alpha C$ at the linear order in $z$. We thus conclude that only the $C$-part of $T_1$ is not conserved. Since the curve is contractible on the minimal surface, this implies that $T_1$ is proportional to $C$, $T_1 = f_1 (\tau) C$. This property follows by induction for all expansion coefficients: Since $\left[ C , \left(\cdot \right) \right] = 0$ and $C^2 = C$, we find that $T_n = f_n (\tau) C$. 

In the following we will show that by subtracting appropriate powers of $T_1$ from each $T_n$ we can construct a charge for which also the $C$-part is conserved. Let us first rewrite \eqn{monocons} as 
\begin{align}
\partial_\tau T[z] = \frac{z}{1-z^2} \left( \partial_\tau T_1 \right) T[z] \, , \label{monocons2}
\end{align}
where we have used that the commutator is always vanishing since $T_n = f_n (\tau) C$. We thus find that $\tilde{T}_2= T_2 - \half T_1 ^2$ is conserved and hence vanishing. Let us now check that for any $T_N$, we can define
\begin{align*}
\tilde{T}_N = T_N - \sum \limits_{M=1}^N  \beta_{N,M} \left(T_1\right)^M
\end{align*}
in such a way, that also the $C$-part of $\tilde{T}_N$ is conserved. We proceed by induction. Suppose that $\partial_\tau \tilde{T}_n = 0$ for all $n < N$. Then we have
\begin{align*}
\partial_\tau T_N = \left(\partial_\tau T_1 \right) \sum \limits _{m=0} ^{N-1} a_m T_m = \left(\partial_\tau T_1 \right)\sum \limits _{m=0} ^{N-1} a_m  \sum \limits _{k=0}^m \beta_{m,k} \left(T_1\right)^k  = \partial_\tau  \sum \limits_{M=1}^N  \beta_{N,M} \left(T_1\right)^M \, .
\end{align*}
The coefficients $a_m$ follow from \eqn{monocons2}, but we need not to know them explicitly. The above argument shows that $\tilde{T}_N$ is indeed conserved and thus vanishing, which concludes the induction. Let us now determine the coefficients $\beta_{N,M}$ explicitly. Equation \eqn{monocons2} can be rewritten as the relation $\partial_\tau T_N - \partial_\tau T_{N-2} = \left(\partial_\tau T_1 \right) T_{N-1}$, which implies that
\begin{align*}
\partial_\tau \Big( \sum \limits_{M=1}^N  \beta_{N,M} \left(T_1\right)^M - \sum \limits_{M=1}^{N-2}  \beta_{N-2,M} \left(T_1\right)^M - \sum \limits_{M=1}^{N-1}  \frac{\beta_{N-1,M}}{M+1} \left(T_1\right)^{M+1} \Big) = 0 \, .
\end{align*}
We hence find the recurrence relation
\begin{align*}
\beta_{N,M} = \beta_{N-2,M} + \frac{1}{M} \beta_{N-1,M-1} \, ,
\end{align*}
which is supplemented by the initial values $\beta_{1,1}=1$, $\beta_{2,2}=\half$, $\beta_{2,1}=0$. The recurrence relation is solved by
\begin{align*}
\beta_{N,N-(2M+1)} = 0 \, , \qquad \beta_{N,N-2M} = \frac{1}{(N-2M)!} \binom{N-1-M}{M}  \, .
\end{align*}
Let us consider the first of the conserved charges $\tilde{T}_N$ explicitly. We note the expansion coefficients
\begin{align}
l_{i, 1} = -2i \, \gamma_{ij} \eps^{jk} J_k \, , \qquad l_{i, 2} = 2 a_i ^{(2)} + \half a_i ^{(1)+(3)} \, ,
\end{align} 
with the Noether current $J_i = g \Lambda_i g^{-1}$. We thus have
\begin{align*}
T_1 &= 2i \, \oint \diff \s \, J_\tau \, , \qquad
T_2 = - 4 \oint \diff \s_1 \diff \s_2 \, \theta(\s_1-\s_2)  J_\tau(\s_1) J_\tau (\s_2) + 2 \oint \diff \s \left( a_\s ^{(2)} + \quarter a_\s ^{(1)+(3)}\right) \, , 
\end{align*}
and defining $\tilde{T}_2 = T_2 - \half T_1 ^2$ gives 
\begin{align*}
\tilde{T}_2 &= - 2 \oint \diff \s_1 \diff \s_2 \, \theta(\s_1-\s_2) \left[ J_\tau(\s_1) , J_\tau (\s_2) \right] + 2 \oint \diff \s \left( a_\s ^{(2)} + \quarter a_\s ^{(1)+(3)}\right) \, . 
\end{align*}
Correspondingly, we have the conserved charges ($\varepsilon(\s) = \theta(\s) - \theta(-\s)$)
\begin{align}
\mathcal{Q}^{(0)} &=  \oint \diff \s \, J_\tau \sim C  \, ,  \\
\mathcal{Q}^{(1)} &=  \frac{1}{2}  \int \diff \s_1 \diff \s_2 \, \varepsilon(\s_1-\s_2) \, \left[ J_\tau(\s_1) \, , \, J_\tau (\s_2) \right] - \oint \diff \s \left( a_\s ^{(2)} + \quarter a_\s ^{(1)+(3)}\right) = 0 \, . \label{charges}
\end{align}

\section{The Level-1 Bonus Symmetry for Super Wilson Loops}
\label{level1B}
In this section, we turn to the strong-coupling description of the super Wilson loop, which has been studied in detail in \cite{Munkler:2015gja}. There, it was shown that the super Wilson loop is invariant under the Yangian $Y[\mathfrak{psu}(2,2 \vert 4)]$ and the level zero and one generators were derived. We extend these results by deriving the precise form of the generator $B^{(1)}$, which is the level-1 recurrence of the hypercharge generator $B$. 

Let us quickly recapitulate the results of \cite{Munkler:2015gja}. The expectation value of the super Wilson loop at strong coupling is given by
\begin{align}
\left \langle \mathcal{W} (\gamma) \right \rangle \overset{\la \gg 1}{=} e^{-\frac{\sqrt{\la}}{2 \pi} \mathcal{A}_\mathrm{ren}(\gamma)} \,.
\end{align}
In this description, $\la$ denotes the 't Hooft coupling constant and  $\mathcal{A}_\mathrm{ren}(\gamma)$ is the minimal area of a superstring ending on the contour $\gamma$ on the conformal boundary superspace. The coset superspace $\mathrm{SU}(2,2\vert4)/(\mathrm{SO}(4,1) \times \mathrm{SO}(5))$ is coordinatized by the coset representatives
\begin{align*}
 g(X,N,y,\theta, \vartheta) = e^{ X \cdot P } \, e^{\theta _\a {} ^A \, Q_A {}^\a + \btheta_{A \da} \, \Qb^{\da A} } \, e^{\vart _A {} ^\a \, S_\a {} ^A + \bvart^{\da A} \, \Sb_{A \da} } \, U(N) \, y^D \, .
\end{align*}
The supermatrix $U(N)$ describes the $\mathrm{S}^5$-part of the superspace, for which we use embedding coordinates $N^I, \, I=1 , \ldots , 6 , \,  N^2=1$. In the above coordinates, the conformal boundary superspace is located at $y=0$ and has half of the fermionic degrees of freedom of the full superspace. We describe the contour $\gamma$ by a parametrization $(x^\mu(\s),\la_\a {} ^A (\s), \bar \la _{A \da}(\s) , n^I(\s)) $ and impose the boundary conditions
\begin{equation}
\begin{alignedat}{3}
X^\mu (\tau = 0 , \s) &= x^\mu (\s) \, , & \qquad \quad 
y(0, \s) &= 0 \, ,  & \qquad \quad 
N^I(0,\s) &= n^I(\s) \, ,  \\
\theta _\a {} ^A (0, \s ) &= \la_\a {} ^A (\s) \, ,  & \qquad \quad
\btheta_{A \da} (0, \s ) &= \bar \la _{A \da} (\s) \, .
\end{alignedat}
\end{equation}
As for the Maldacena-Wilson loop, the minimal area is divergent and the finite area $\mathcal{A}_\mathrm{ren}(\gamma)$ is computed by introducing a cut-off $\varepsilon$ in the $y$-direction and subtracting the divergence,
\begin{align}
	\mathcal{A}_\mathrm{ren}(\gamma) = \lim \limits_{\varepsilon \to 0} \left \lbrace
		\mathcal{A}_{\mathrm{min}}(\gamma) \Big \vert _{y \geq \varepsilon}
		- \frac{\mathcal{L}(\gamma)}{\varepsilon}
	\right \rbrace \, ,
	&& \mathcal{L}(\gamma) = \int \diff \s \, \lvert \pi(\s) \rvert \, .
\end{align}
Here, $\pi ^\mu = \dx ^\mu + i \big( \dot{\bar{\la}} \s ^\mu \la - \bar{\la} \s ^\mu \dot{\la} \big)$ is the supermomentum of a particle moving along the contour $\gamma$. Solving the equations of motion iteratively in an expansion in $\tau$ allows to derive the first few coefficients of the parametrization of the minimal surface. More precisely, using the notation $F(\tau,s)= \sum F_{(n)}(s) \tau ^n$, we have
\begin{align}
X_{(1)} = 0 \,, \qquad \theta_{(1)} = 0 \, , \qquad \btheta_{(1)} = 0 \, , \qquad \vart _{(0)} {} _A {}^\a =  i \,  \dot{\bar{\la}}_{A \da} \, \pi^{\da \a} \, , \qquad \bvart _{(0)} {}^{\da A} = - i \, \pi ^{\da \a} \, \dot{\la} _\a {}^A \,. 
\end{align}
Also the coefficients $X_{(2)}, \theta_{(2)}$ can be derived in terms of the boundary data, they are given by
\begin{align}
X_{(2)} ^\mu &= \dot{\pi}^\mu + i \tr \big( \btheta_{(2)} \s^\mu \la - \bla \s^\mu \theta_{(2)} \big) \, , \\
\theta_{(2)} {} _\a {} ^A &= - \dot{\pi}_{\a \da} \, \pi ^{\da \b} \, \dot{\la}  _\b {} ^A + \bar K^{\prime \, A B} \left( 4 \, \big( \dot{\bar{\la}} \pi \dot{\la} \big)_B {} ^C  - i \,  \delta ^C _B \, \partial _s \right) \big( \pi _{\a \db} \, \dot{\bar{\la}}  _C {} ^\db \big) \label{theta2} \, , \\
\btheta _{(2) \, A \da} &= - \, \dbla _{A \db} \, \pi ^{\db \a} \, \dot{\pi}_{\a \da} -  K^\prime _{A B} \left( 4 \, \big( \dbla\pi \dot{\la} \big)_C {} ^B  + i \,  \delta ^B _C \, \partial _s \right) \big( \dot{\la}^{\b C} \, \pi_{\b \da} \big) \, .
\end{align}
The matrices $\bar K^{\prime \, A B}$ and $K^\prime _{A B}$ are given by equation \eqn{Kdef} and
\begin{align}
K^\prime _{A B} = u _A {} ^C \, u _B {} ^D \, K_{CD} = \left( u K u^T \right)_{AB} \, , \qquad 
\bar K^{\prime \, A B} = \left( \left(u^{-1} \right)^T K u^{-1} \right) ^{AB} \, .
\end{align}
Here, $u _A {} ^C$ denote the entries in the $\mathrm{SU}(4)$-part of the supermatrix $U(N)$, which we use to describe the $\mathrm{S}^5$-part of the coset space. The higher-order coefficients $X_{(3)}, \theta_{(3)}, \vart_{(1)}, N_{(1)}$ are related to functional derivatives of the minimal area $\mathcal{A}_\mathrm{ren}(\gamma)$, see \cite{Munkler:2015gja} for explicit formulae. Given these results, one can evaluate the charges \eqn{charges} on the minimal surface. For the level-zero charge $\mathcal{Q}^{(0)} = \mathcal{Q}^{(0)}_a \hat{T}^a$, we have
\begin{align*}
\mathcal{Q}^{(0)}_a  = \int \diff \s j_a (\s) (\mathcal{A}_\mathrm{ren}(\gamma)) \, , 
\end{align*}
where the densities $j_a(\s)$ are provided in appendix \ref{app:generators}. They form a representation of $\mathfrak{psu}(2,2 \vert 4)$ if we restrict to the respective conserved charges. The vanishing of these charges thus encodes the superconformal symmetry of the super Wilson loop at strong coupling, 
\begin{align*}
J_a ^{(0)} \langle \mathcal{W}(\gamma) \rangle \overset{\la \gg 1}{=} - \frac{\sqrt{\la}}{2 \pi} \, \mathcal{Q}^{(0)}_a \langle \mathcal{W}(\gamma) \rangle  = 0 \, , \quad \text{where} \quad  J_a ^{(0)} = \int \diff \s j_a ( \s ) \, .
\end{align*}
The coefficients $\mathcal{Q}^{(0)}_a$ are contracted with the generators $\hat{T}^a = G^{ab} T_b $ of the dual basis, which can be defined for $\mathfrak{psu}(2,2\vert4)$ or $\mathfrak{u}(2,2\vert4)$, for which the metric $G_{ab}$ is non-degenerate. In order to discuss the $C$-part of $\mathcal{Q}^{(0)}$, we need to formally enlarge the algebra to $\mathfrak{u}(2,2\vert 4)$, where we have the scalar product $\langle B , C \rangle =1$. Then we see that the $C$-part of $\mathcal{Q}^{(0)}$ leads to the hypercharge generator 
\begin{align}
B^{(0)} = \frac{1}{2} \int \diff \s \left( \la _\a {} ^A (\s) \, \frac{\delta}{\delta \la _\a {} ^A (\s)} - \bla _{A \da} (\s) \, \frac{\delta}{\delta \bla _{A \da}  (\s)}  \right) \, , 
\end{align}   
which does not provide a symmetry of the super Wilson Loop as the $C$-part of $\mathcal{Q}^{(0)}$ does not vanish.

The evaluation of the level-one charge $\mathcal{Q}^{(1)}$ allows to read off the level-one Yangian generators from the relation
\begin{align}
J_a ^{(1)} \langle \mathcal{W}(\gamma) \rangle \overset{\la \gg 1}{=} \left( \frac{\la}{2 \pi^2} \, \mathcal{Q}^{(1)}_a + \mathcal{O}\big( \sqrt{\la} \big) \right) \langle \mathcal{W}(\gamma) \rangle \, .
\end{align}
The level-1 recurrences of the $\mathfrak{psu}(2,2\vert 4)$-generators have been constructed in \cite{Munkler:2015gja}. Here, we focus on the level-1 hypercharge $B^{(1)}$, which does provide a symmetry of the super Wilson loop since also the C-part of $\mathcal{Q}^{(1)}$ vanishes. We find 
\begin{align}
B^{(1)} & = \int \diff \s_1 \diff \s_2 \, \varepsilon(\s_1-\s_2) \ft{1}{8} \left( q_A ^\a (\s_1) s_\a ^A (\s_2) + \bar{q} ^{A \da} (\s_1) \bar{s}_{\da A} (\s_2) + s_\a ^A (\s_1) q_A ^\a (\s_2)  + \bar{s}_{\da A} (\s_1)  \bar{q} ^{A \da} (\s_2)  \right) \nn \\
& - \int \diff \s \, \tr \left[ \la \bla \pi \right] \left( \delta^{IJ} - n^I n^J \right) \frac{\delta^2}{\delta n^I(\s) \delta n^J (\s) }  \nn \\
& + \frac{\la}{2 \pi^2} \int \diff \s \Big \lbrace 3 \tr \left[ \big(\dot{\la}\bla - \la \dot{\bla} \big) \pi \dot{\la} \dot{\bla} \pi    + 2 \la \bla \big(\dot{\pi} \dot{\la} \dot{\bla} \pi - \pi \dot{\la} \dot{\bla} \dot{\pi} \big) \right] + 8 \tr \left[\la \bla \eps \big( \dot{\la} \rho \dot{\bla} \big)^T \eps   \right] \label{B1} \\
& \qquad \quad + 2 i \tr \left[ \big( \dot{\la} \dot{\bla} + \la \rho \dot{\bla} - \dot{\la} \rho \bla +\la \btheta_{(2)} + \theta_{(2)} \bla   \big) \pi \right] - i \tr \left[ \la \bla \pi \right] \left( \dot{n}^2 - \dot{\pi}^2 \right)  \nn \\
& \qquad \quad + i \tr \left[ \la \bla \pi \right] \tr \left[\big( 12 \dot{\la} \dot{\bla} \pi \dot{\la} \dot{\bla}  + 2i \big( \dot{\la} \ddot{\bla} - \ddot{\la} \dot{\bla} \big) + 4i \big( \theta_{(2)} \dot{\bla} - \dot{\la} \btheta_{(2)} + \dot{\la} \rho \dot{\bla} \big) \big) \pi \right]  \Big \rbrace \nn \, . 
\end{align}
To shorten the local term, we used the abbreviations
\begin{align*}
\rho_A {} ^B =   n^I \dot{n} ^J \, \left(\gamma ^{IJ} \right) _A {} ^B -  i \big( \dot{\bla}  & \pi \dot{\la}  \big)_A {} ^B  + i \bar K^{\prime \, BC}  \big( \dot{\bla} \pi \dot{\la} \big)_C {} ^D  K^\prime _{DA} - \ihalf \delta ^B _A  \tr \left[ \dot{\la} \dot{\bla} \pi  \right] \, , \qquad \eps = \begin{pmatrix}
0 & -1 \\ 1 & 0
\end{pmatrix} \, .
\end{align*}
Here, the matrices $\gamma^{IJ}$ span $\mathfrak{su}(4)$, see appendix B of \cite{Munkler:2015gja} for details. The higher-level hypercharge generators $B^{(N)}$ can be obtained in a similar way by computing the $C$-part of the conserved charges $\tilde{T}_N$. From the structure of the charges $\tilde{T}_N$ it is clear that the generators derived in this way will contain an $(N+1)$-parameter ordered integral involving functional variations at each point. For the moment, we focus on the level-1 hypercharge generator. 

The bi-local part in the first line of equation \eqn{B1} shows the typical structure of the level-1 hypercharge as it has been observed in other cases as well. The double-functional derivative in the second line appears for all level-one generators in a similar fashion. In the strong-coupling limit it reduces to the product of functional derivatives of $\mathcal{A}_\mathrm{ren}(C)$ and so the double functional derivative at the same point along the loop does not give rise to a divergence. One would expect that this part of the generator requires a point-splitting regularization at weak coupling and it should be illuminating to compare this part of the generator with the weak-coupling Yangian symmetry generators of \cite{Beisert:2015xxx}. This is also the case for the last piece of the generator, which does not involve functional derivatives and hence simply integrates to a number, although it depends on the boundary curve in a complicated fashion. 

At first sight, the bi-local part of the above generator seems to depend on the choice of a starting point along the contour due to the path-ordering prescription that is encoded in the antisymmetric step function. The bi-local part of any level-one generator $J_a^{(1)}$ is given by the ordered integral
\begin{align*}
f \indices{^{cb} _a}  \int \diff \s_1 \diff \s_2 \, \varepsilon(\s_1-\s_2) \, j_b(\s_1) \, j_c(\s_2) \, ,
\end{align*}
where $j_b(\s)$ are the variational densities provided in appendix \ref{app:generators} and $f \indices{^{cb} _a}$ denote the structure constants of $\mathfrak{u}(2,2\vert4)$ in the basis given by the $j_a$. If instead of $x(0)$ one chooses $x(\Delta)$ as the starting point for the ordered integral, the difference between two level-one generators is given by
\begin{align*}
J_a^{(1)} - \tilde{J}_a ^{(1)} = f\indices{^{cb}_a} f\indices{_{bc}^d} \, \int \limits _0 ^\Delta  j_d (\s)
\end{align*}
If $J_a^{(1)}$ is any of the level-one generators of $Y[\mathfrak{psu}(2,2\vert4)]$, this difference is vanishing because the contraction $f\indices{^{cb}_a} f\indices{_{bc}^d}$ vanishes due to the vanishing of the dual Coxeter number of $\mathfrak{psu}(2,2\vert4)$. For the level-1 hypercharge generator one finds that 
\begin{align*}
B^{(1)} - \tilde{B}^{(1)} \sim \int \limits _0 ^\Delta c(\s) \, , 
\end{align*}
which vanishes as $c(\s)$ vanishes identically in our superspace representation of $\mathfrak{u}(2,2\vert 4)$.

\section{Conclusion and Outlook}
\label{sec:outlook}
In this paper we have shown that the super Wilson loop in $\mathcal{N}=4$ SYM is invariant under the so-called bonus symmetry, which is the level-1 recurrence of the hypercharge generator. For this generator, there is no analogous symmetry at level zero, as the hypercharge generator is itself not a symmetry of the super Wilson loop. This finding enlarges the Yangian $Y[\mathfrak{psu}(2,2\vert4)]$ symmetries of the super Wilson loop, which thus provides a new observable that is invariant under the bonus symmetry. 

The derivation relies on the construction of new conserved charges in the central extension of the supercoset space of type IIB superstring theory on $\mathrm{AdS}_5 \times \mathrm{S}^5$. A similar construction has been performed for the pure spinor superstring \cite{Berkovits:2011kn}, where new conserved charges were constructed in all odd levels. Here, the disk topology of the minimal surface allows for a simple construction, which contains charges in all levels except level zero.

It is interesting to note that, given the superconformal symmetry of the super Wilson loop, all other level-one generators can be inferred from $B^{(1)}$ by repeated application of the commutation relation
\begin{align*}
\big[ J_a ^{(0)} , J_b^{(1)} \big \rbrace = f_{ab} {} ^c J_c^{(1)} \, .
\end{align*}
In contrast, the level-1 hypercharge generator $B^{(1)}$ cannot be generated from the other level-one generators in this way. The situation is similar in the higher levels, where the hypercharge recurrences cannot be constructed from commutators of lower-level generators. The conserved charges obtained in section \ref{sec:charges} point to the existence of a hypercharge generator in any higher level and, as all these new generators should be algebraically independent of the other symmetry generators, they provide a large class of new symmetries for the super Wilson loop. However, it should be noted that it has not yet been verified that these generators form a Yangian symmetry algebra as the Serre relations have not been checked. Moreover, the authors of \cite{deLeeuw:2012jf} have noted algebraic obstructions\footnote{I thank Florian Loebbert for drawing my attention to this point.} against the existence of hypercharge-like generators in the even levels. For this question, the superstring calculation presented in this paper could be interesting as it in principle allows to derive candidates for these generators and to check their algebraic relations with the other generators.

\subsection*{Acknowledgements}
I would like to thank Florian Loebbert, Dennis M\"uller and Jan Plefka for interesting discussions as well as Florian Loebbert, Jan Plefka and Konstantin Zarembo for valuable comments on the draft.

This research is supported by the SFB 647 \textit{``Raum-Zeit-Materie.
Analytische und Geometrische Strukturen''} and the Research Training Group GK 1504
\textit{``Mass, Spectrum, Symmetry''}.

\newpage
\appendix

\section{The Fundamental Representation of \texorpdfstring{$\mathfrak{u}(2,2\vert4)$}{u(2,2;4)}}
\label{app:u224}
In this appendix, we provide our conventions for the fundamental representations of $\mathfrak{u}(2,2 \vert 4)$, which follow \cite{Drummond:2009fd}. A more detailed exposition can be found in the review \cite{Arutyunov:2009ga}. The superalgebra $\mathfrak{u}(2,2\vert 4)$ can be defined as the set of $(4\vert 4)$ supermatrices satisfying the following reality condition:
\begin{align}
B = \begin{pmatrix}
m & \theta \\ \eta & n
\end{pmatrix} = \begin{pmatrix}
-H\, m^\dag H^{-1} & -H\, \eta^\dag   \\ -\theta^\dag H^{-1} & - n ^\dag
\end{pmatrix} = - \begin{pmatrix}
H & 0 \\ 0 & \mathbb{I}_4 
\end{pmatrix} B^\dag \begin{pmatrix}
H^{-1} & 0 \\ 0 & \mathbb{I}_4 
\end{pmatrix} \,.
\end{align}
Here, the matrix $H$ is given by
\begin{align}
H = \begin{pmatrix}
0 & \mathbb{I}_2  \\ \mathbb{I}_2 & 0  
\end{pmatrix} \,.
\end{align}
To endow $\mathfrak{u}(2,2\vert4)$ with a $\mathbb{Z}_4$-grading consider the following automorphism of $\mathfrak{gl}(4 \vert 4)$:
\begin{align}
B \mapsto \Omega(B) = - \mathcal{K} \, B^{\mathrm{st}} \, \mathcal{K}^{-1} \, , \qquad \mathcal{K} = \begin{pmatrix}
K & 0 \\ 0 & K
\end{pmatrix} \,, \qquad K = \begin{pmatrix}
-i \s^2 & 0 \\ 0 & -i \s^2
\end{pmatrix} \, . \label{Kdef}
\end{align}
Based on this automorphism one can define a projection operator by
\begin{align}
P^{(k)}(B) =  B^{(k)} = \quarter \left( B + i^{3k} \Omega(B) +  i^{2k} \Omega^2(B) +  i^{k} \Omega^3(B) \right) \, .
\end{align}
A grading on $\mathfrak{u}(2,2\vert4)$ can then be defined by
\begin{align*}
\mathfrak{u}(2,2\vert4) &= \mathfrak{g}^{(0)} \oplus \mathfrak{g}^{(2)} \oplus \mathfrak{g}^{(1)} \oplus \mathfrak{g}^{(3)} \, , \quad \text{where} \quad \mathfrak{g}^{(k)} := \left\lbrace  P^{(k)}(B) \, \vert \, B \in \mathfrak{u}(2,2\vert4) \right \rbrace \, , \\
\left[ \mathfrak{g}^{(k)} \, , \, \mathfrak{g}^{(l)} \right] &\subset \mathfrak{g}^{(k+l) \, \mathrm{mod} 4 }  \,.
\end{align*}
We choose the following basis for the superalgebra $\mathfrak{u}(2,2\vert 4)$: 
\begin{align}
\left( \begin{array}{cc|c} 0 & P_\mu & Q _A {} ^\a \\
K_\mu & 0 & \bar{S}_{A \da} \\ \hline
S _\a {} ^A & \bar{Q}^{\da A} & R \indices{^A _B}
\end{array} \right) = 
\left( \begin{array}{cc|c} 0 & i \bs_\mu & 2 \, E \indices{^\a _A} \\
i \s_\mu & 0 & 2 \, E_{\da A} \\ \hline
- 2 \, E \indices{^A _\a} & -2 \, E^{A \da} & 4 \, E \indices{^A _B} - \delta ^A _B \, \mathbb{I}_4
\end{array} \right) 
\end{align}
This equation is to be read as
\begin{align}
P_\mu = \left( \begin{array}{cc|c} 0 & i \bs_\mu & 0 \\
0 & 0 & 0 \\ \hline
0 & 0 & 0
\end{array} \right)
\end{align}
and similarly for the other generators. The notation $E \indices{^A _B}$ denotes a matrix with entry 1 in the position $(A,B)$ and all other entries vanishing. The remaining generators of  $\mathfrak{u}(2,2\vert4)$ are given by
\begin{align}
M_{\mu \nu}  = -\frac{i}{2} \left( \begin{array}{c|c}
		 \begin{array}{cc}
		\s_{\mu \nu} & 0 \\ 0 & \bs_{\mu \nu} 
\end{array}	& 0 \\
		\hline
		0 & 0 
	\end{array} \right) \, , &&
C  = \frac{1}{2} \left( \begin{array}{c|c}
		 \mathbb{I}_4 & 0 \\
		\hline
		0 & \mathbb{I}_4  
	\end{array} \right) \, ,  \\
D  = \frac{1}{2} \left( \begin{array}{c|c}
		 \begin{array}{cc}
		\mathbb{I}_2  & 0 \\ 0 & -\mathbb{I}_2 
\end{array}	& 0 \\
		\hline
		0 & 0 
	\end{array} \right) \, , &&
B  = -\frac{1}{2} \left( \begin{array}{c|c}
		 0 & 0 \\
		\hline
		0 & \mathbb{I}_4  
	\end{array} \right) \, .
\end{align}
We provide the commutation relations for the above generators. The commutators with the generators $M$ and $R$ only depend on the set of indices and their position:
\begin{equation}
\begin{alignedat}{2}
\Big[ M \indices{_\a ^\b} \, , \, J_\g \Big] &= 2i \, \delta ^\b _\g J_\a - i \delta ^\b _\a J_\g \, & \qquad \qquad  \Big[ M \indices{_\a ^\b} \, , \, J^\g \Big] &= - 2i \, \delta ^\g _\a J^\b + i \delta ^\b _\a J^\g \\
\Big[ \overline{M} \, \indices{^\da _\db} \, , \, J^\dg \Big] &= 2i \, \delta ^\dg _\db J^\da - i \delta ^\da _\db J^\dg \, & \qquad \qquad  \Big[ \overline{M} \, \indices{^\da _\db} \, , \, J_\dg \Big] &= - 2i \, \delta ^\da _\dg J_\db + i \delta ^\da _\db J_\dg \\
\Big[ R \indices{^A _B} \, , \, J^C \Big] &= 4 \, \delta ^C _B J^A - \delta ^A _B J^C \, & \qquad \qquad \Big[ R \indices{^A _B} \, , \, J_C \Big] &= - 4 \, \delta ^A _C J_B + \delta ^A _B J_C
\end{alignedat}
\end{equation}
The commutators with the dilatation $D$ and hypercharge generator $B$ are specified by a weight $\Delta(T_a)$ or a hypercharge $\mathrm{hyp}(T_a)$,
\begin{align*}
\left[ D, T_a \right] = \Delta(T_a) \, T_a \, , \qquad \left[ B, T_a \right] = \mathrm{hyp}(T_a) \, T_a \, .
\end{align*}
The non-vanishing weights or hypercharges of the generators are given by
\begin{align}
\begin{aligned}
\Delta(P) = 1 \, ,\qquad  && \Delta(Q, \Qb) = \half \, ,\qquad  && \Delta(S, \Sb)= - \half \, , \\
\Delta(K)=-1 \, ,\qquad  && \mathrm{hyp}(Q,\Sb) = \half \, , \qquad && \mathrm{hyp}(\Qb,S) =- \half \, .
\end{aligned}
\end{align}
Moreover, we note the following commutation relations:
\begin{equation}
\begin{alignedat}{3}
\Big[ K_{\a \da} \, , \, Q_A {}^\b \Big] &= -2i \, \delta^\b _\a \bar{S}_{A \da} \, & \quad \Big[ K_{\a \da} \, , \, \bar{Q}^{\db A} \Big] &= + 2i \, \delta^\db _\da S _\a {}^A \, & \quad  \Big \lbrace Q _A {}^\a \, , \, \bar{Q}^{\da B} \Big \rbrace &= -2i \, \delta ^B _A P^{\da \a} \\
\Big[ P^{\da \a} \, , \, S_\b {} ^A \Big] &= + 2i \, \delta^\a _\b \bar{Q}^{\da A} \, & \quad \Big[ P^{\da \a} \, , \, \bar{S}_{A \db} \Big] &= - 2i \, \delta^\da _\db Q_A {}^\a \, & \quad \Big \lbrace S _\a {} ^A \, , \, \bar{S}_{B \da} \Big \rbrace &= -2i \, \delta ^A _B K_{\a \da}
\end{alignedat}
\end{equation}
The remaining non-vanishing commutators are given by
\begin{equation}
\begin{aligned}
\Big[ P_\mu \, , \, K_\nu \Big] &= + 2 \eta _{\mu \nu} \, D - 2 M_{\mu \nu} \, , \\
\Big \lbrace Q _A {} ^\a \, , \, S_\b {}^B \Big \rbrace &= -2i \, \delta ^B _A \, M \indices{_\b ^\a} - \delta ^\a _\b \, R \indices{^B _A} - 2 \, \delta ^B _A \, \delta ^\a _\b \left(D + C \right) \, ,\\
\Big \lbrace \bar{Q}^{\da A} \, , \, \bar{S}_{B \db} \Big \rbrace &= -2i \, \delta ^A _B \, \overline{M} \, \indices{^\da _\db} - \delta ^\da _\db \, R \indices{^A _B} + 2 \, \delta ^A _B \, \delta ^\da _\db \left(D - C \right) \,.
\end{aligned}
\end{equation}
We collectively denote the generators defined above by $T_a$ and their structure constants by $\tilde{f}_{ab} {} ^c$, 
\begin{align}
\Big[ T_a \, , \, T_b \Big \rbrace = \tilde{f}_{ab} {} ^c \, T_c \,.
\end{align}
The metric $\tilde{G}_{a b} = \left \langle T_a  ,  T_b \right \rangle = \str (T_a T_b)$ on the algebra has the following components:
\begin{equation}
\begin{alignedat}{3}
\langle P^{\da \a} , K_{\b \db} \rangle &= - 4 \, \delta ^\a _\b \, \delta ^\da _\db & \quad \langle M \indices{_\a ^\b} , M \indices{_\g ^\eps} \rangle &= - 4 \, \delta ^\b _\g \, \delta ^\eps _\a + 2\, \delta ^\b _\a \, \delta ^\eps _\g & \quad  \langle D , D \rangle &= 1 \\
\langle Q _A {} ^\a , S_\b {} ^B \rangle &= -4 \, \delta ^B _A \, \delta ^\a _\b \, \delta ^\da _\db    & \quad  \langle \bar{M} \indices{^\da _\db} , \bar{M} \indices{^\dg _{\dot{\eps}}} \rangle &= - 4 \, \delta ^\dg _\db \, \delta ^\da _{\dot{\eps}} + 2\, \delta ^\da _\db \, \delta ^\dg _{\dot{\eps}} & \quad  \langle B , C \rangle &= 1   \\
\langle \bar{Q}^{\da A} , \bar{S}_{B \db} \rangle &=  4 \, \delta ^A _B \, \delta ^\da _\db   & \quad  \langle R \indices{^A _B} , R \indices{^C _D} \rangle &= - 16 \, \delta ^A _D \, \delta ^C _B + 4 \, \delta ^A _B \, \delta ^C _D  \label{groupmetric}
\end{alignedat}
\end{equation}
All other entries are vanishing. Note that $\tilde{G}_{a b}$ satisfies the symmetry property $\tilde{G}_{a b} = \left(-1 \right) ^{\lvert a \rvert} \tilde{G}_{b a}$, where $\lvert a \rvert = \mathrm{deg}(T_a)$ denotes the Gra{\ss}mann degree of a (homogeneous) basis element, $\lvert a \rvert = 0 \, (1)$ for an even (odd) generator.

\section{Superspace Representation of \texorpdfstring{$\mathfrak{u}(2,2\vert4)$}{u(2,2|4)}}
\label{app:generators}
In this appendix we provide the differential generators $j_a (\s)$ obtained from
\begin{align}
j_a (\s) \left( \mathcal{A} \right) = \left\langle J_{\tau \, (0)}(\s) \, , \, T_a \right\rangle \, ,
\end{align}
which we write out explicitly in the form $p^\mu (\s) \left( \mathcal{A} \right) = \left\langle J_{\tau \, (0)} (\s)\, , \, P^\mu \right\rangle$ and similarly for all other generators. We use the short-hand notation
\begin{align}
\partial^\mu = \frac{\delta }{\delta x_\mu (\s)} \, , \quad \partial^{\da \a} = \sigma _\mu ^{\da \a} \, \partial ^\mu  \, , \quad \partial ^\a _A = \frac{\delta }{\delta \la _\a ^A(\s)} \, , \quad \bar{\partial} ^{A \da} =  \frac{\delta }{\delta \bar{\la}_{A \da}(\s)} \, , \quad \partial ^I = \frac{\delta }{\delta n^I (\s)} \,.
\end{align}
Then, we have:
\begin{equation}
\begin{alignedat}{2}
p^{\mu} &= \partial^{\mu}  \, & \qquad  d &= \half \left( \la \, \partial _\la + \bar{\la} \, \bar{\partial} _{\la} \right) + x \cdot \partial _x \\
q^\a _A &= - \partial ^\a _A + i \bar{\la}_{A \da} \, \partial ^{\da \a} \, & \qquad  \bar{q}^{A \da} &= - \bar{\partial}^{A \da} + i \la ^A _\a \, \partial ^{\da \a} \label{levelzero1}  \\
m \indices {_\a ^\b} &= n \indices {_\a ^\b} - \half \, \delta ^\b _\a \, n \indices {_\g ^\g} \, & \qquad  \bar{m} \indices{ ^\da _\db} &= \bar{n} \indices{ ^\da _\db} - \half \delta ^\da _\db \, \bar{n} \indices{ ^\dg _\dg} \\
n \indices {_\a ^\b} &= -2i \, \la ^A _\a \, \partial ^\b _A + i \, x_{\a \da} \, \partial ^{\da \b} & \qquad  \bar{n} \indices{ ^\da _\db} &= 2i \, \bar{\la}_{A \db} \, \bar{\partial}^{A \da} - i \, x_{\a \db} \, \partial ^{\da \a} \\
b&= \half \left( \la \, \partial _\la - \bar{\la} \, \bar{\partial} _{\la} \right) & \qquad c&= 0   
\end{alignedat}
\end{equation}
The remaining generators are given by:
\begin{equation}
\begin{aligned}
 r \indices{^A _B} &= 4 \big( \big( \gamma ^{IJ} \big) _B {} ^A  \, n^I \partial ^J  + \bla _{B \da} \, \bar{\partial} ^{A \da}  -  \la ^A _\a \, \partial ^\a _B  \big) -   \delta ^A _B  \big(  \bla \, \bar{\partial}_\la  - \la \,  \partial _\la \big) \\
s^A _\a &= i \, x^{-} _{\a \da} \, \bar{\partial}^{A \da} + x^{+} _{\a \da} \, \la ^A _\b \, \partial ^{\da \b} - 4 \, \la ^B _\a \, \la ^A _\b \, \partial^\b _B + 4 \la ^B _\a \, \big( \gamma ^{IJ} \big) _B {} ^A \, n^I \partial ^J
\\
\bar{s}_{A \da} &= -i \, x^{+} _{\a \da} \, \partial ^\a _A - x^{-} _{\a \da}  \, \bla_{A \db} \, \partial ^{\db \a} - 4 \, \bla _{A \db} \, \bla _{B \da} \bar{\partial}^{B \db} + 4 \big( \gamma ^{IJ} \big) _A {} ^B \, \bla _{B \da} \, n^I \partial ^J
\\
k_{\a \da} &= i \, x^+ _{\a \db} \, \bar{n}^\db {} _\da - i x^- _{\b \da} \, n _\a {} ^\b - x^+_{\a \db} \, x^- _{\b \da} \, \partial ^{\db \b} - 8i \left( \la \, \gamma ^{IJ} \, \bla \right)_{\a \da} \, n^I \partial ^J   \label{levelzero2} 
\end{aligned}
\end{equation}
Here, we introduced the chiral and anti-chiral coordinates
\begin{align}
x^+ _{\a \da} = x_{\a \da} + 2i \, \la _\a {}^A \, \bla _{A \da} \, , \qquad \qquad x^- _{\a \da} = x_{\a \da} - 2i \, \la _\a {}^A \, \bla _{A \da} \,.
\end{align}  
A comparison to the commutation relations of the generators given in appendix~\ref{app:u224} shows that all except the odd-odd commutators have a different sign,
\begin{equation}
\begin{aligned}
\Big[ j_a (\s) \, , \, j_b (\s^\prime) \Big \rbrace &= f \indices{_{ab} ^c}  \, \delta(\s - \s^\prime) \, j_c (\s) \, , \qquad f \indices{_{ab} ^c} = \tilde{f} \indices{_{ba} ^c} = - (-1)^{ \lvert a \rvert \lvert b \rvert}  \tilde{f} \indices{_{ab} ^c} 
\end{aligned}
\end{equation}

\newpage

%%%%%%%%%%%%%%%%%%%%%%%%%%%%%%%%%%%%%%%%%%%%%%%%%%%%%%%%%%%%%%%%%%%%%%%%%%%%%%%%
\bibliographystyle{nb}
\bibliography{botany}

\begin{thebibliography}{10}
\ifx\href\asklfhas\newcommand{\href}[2]{#2}\fi
\ifx\arxivref\asklfhas\newcommand{\arxivref}[2]{\href{http://arxiv.org/abs/#1}{#2}}\fi
\ifx\doiref\asklfhas\newcommand{\doiref}[2]{\href{http://dx.doi.org/#1}{#2}}\fi
\raggedright
\small
\parskip 0pt

%\%CITATION = 0902.2987;\%\%
\bibitem{Drummond:2009fd}
J.~M.~Drummond, J.~M.~Henn and J.~Plefka,
\textit{``{Yangian symmetry of scattering amplitudes in
  {$\mathcal{N}=\mathord{}$4} super Yang-Mills theory}''},
\textsf{\doiref{10.1088/1126-6708/2009/05/046}{JHEP~0905,~046~(2009)}},
\texttt{\arxivref{0902.2987}{arxiv:0902.2987}}.

%\%CITATION = ARXIV:1002.1733;\%\%
\bibitem{Beisert:2010gn}
N.~Beisert, J.~Henn, T.~McLoughlin and J.~Plefka,
\textit{``{One-Loop Superconformal and Yangian Symmetries of Scattering
  Amplitudes in N=4 Super Yang-Mills}''},
\textsf{\doiref{10.1007/JHEP04(2010)085}{JHEP~1004,~085~(2010)}},
\texttt{\arxivref{1002.1733}{arxiv:1002.1733}}.

%\%CITATION = ARXIV:0908.2437;\%\%
\bibitem{Sever:2009aa}
A.~Sever and P.~Vieira,
\textit{``{Symmetries of the N=4 SYM S-matrix}''},
\texttt{\arxivref{0908.2437}{arxiv:0908.2437}}.

%\%CITATION = ARXIV:0708.1285;\%\%
\bibitem{Matsumoto:2007rh}
T.~Matsumoto, S.~Moriyama and A.~Torrielli,
\textit{``{A Secret Symmetry of the AdS/CFT S-matrix}''},
\textsf{\doiref{10.1088/1126-6708/2007/09/099}{JHEP~0709,~099~(2007)}},
\texttt{\arxivref{0708.1285}{arxiv:0708.1285}}.

%\%CITATION = ARXIV:1103.0646;\%\%
\bibitem{Beisert:2011pn}
N.~Beisert and B.~U.~Schwab,
\textit{``{Bonus Yangian Symmetry for the Planar S-Matrix of N=4 Super
  Yang-Mills}''},
\textsf{\doiref{10.1103/PhysRevLett.106.231602}{Phys.Rev.Lett.~106,~231602~(2011)}},
\texttt{\arxivref{1103.0646}{arxiv:1103.0646}}.

%\%CITATION = ARXIV:1106.2536;\%\%
\bibitem{Berkovits:2011kn}
N.~Berkovits and A.~Mikhailov,
\textit{``{Nonlocal Charges for Bonus Yangian Symmetries of
  Super-Yang-Mills}''},
\textsf{\doiref{10.1007/JHEP07(2011)125}{JHEP~1107,~125~(2011)}},
\texttt{\arxivref{1106.2536}{arxiv:1106.2536}}.

%\%CITATION = HEP-TH/9803002;\%\%
\bibitem{Maldacena:1998im}
J.~M.~Maldacena,
\textit{``{Wilson loops in large N field theories}''},
\textsf{\doiref{10.1103/PhysRevLett.80.4859}{Phys.Rev.Lett.~80,~4859~(1998)}},
\texttt{\arxivref{hep-th/9803002}{hep-th/9803002}}.

%\%CITATION = HEP-TH/0007104;\%\%
\bibitem{Ooguri:2000ps}
H.~Ooguri, J.~Rahmfeld, H.~Robins and J.~Tannenhauser,
\textit{``{Holography in superspace}''},
\textsf{\doiref{10.1088/1126-6708/2000/07/045}{JHEP~0007,~045~(2000)}},
\texttt{\arxivref{hep-th/0007104}{hep-th/0007104}}.

%\%CITATION = ARXIV:1506.07047;\%\%
\bibitem{Beisert:2015jxa}
N.~Beisert, D.~Muller, J.~Plefka and C.~Vergu,
\textit{``{Smooth Wilson Loops in N=4 Non-Chiral Superspace}''},
\texttt{\arxivref{1506.07047}{arxiv:1506.07047}}.

%\%CITATION = ARXIV:1309.1676;\%\%
\bibitem{Muller:2013rta}
D.~M{\"u}ller, H.~M{\"u}nkler, J.~Plefka, J.~Pollok and K.~Zarembo,
\textit{``{Yangian Symmetry of smooth Wilson Loops in $\mathcal{N} = $ 4 super
  Yang-Mills Theory}''},
\textsf{\doiref{10.1007/JHEP11(2013)081}{JHEP~1311,~081~(2013)}},
\texttt{\arxivref{1309.1676}{arxiv:1309.1676}}.

\bibitem{Beisert:2015xxx}
N.~Beisert, D.~M{\"u}ller, J.~Plefka and C.~Vergu,
\textit{``{Work in Progress}''}.

%\%CITATION = ARXIV:1503.07553;\%\%
\bibitem{Munkler:2015gja}
H.~M{\"u}nkler and J.~Pollok,
\textit{``{Minimal Surfaces of the $AdS_5\times S^5$ Superstring and the
  Symmetries of Super Wilson Loops at Strong Coupling}''},
\texttt{\arxivref{1503.07553}{arxiv:1503.07553}}.

%\%CITATION = 0901.4937;\%\%
\bibitem{Arutyunov:2009ga}
G.~Arutyunov and S.~Frolov,
\textit{``{Foundations of the $AdS_5 \times S^5$ Superstring. Part I}''},
\textsf{\doiref{10.1088/1751-8113/42/25/254003}{J.~Phys.~A42,~254003~(2009)}},
\texttt{\arxivref{0901.4937}{arxiv:0901.4937}}.

%\%CITATION = HEP-TH 0305116;\%\%
\bibitem{Bena:2003wd}
I.~Bena, J.~Polchinski and R.~Roiban,
\textit{``{Hidden symmetries of the {$AdS_5\times S^5$} superstring}''},
\textsf{\doiref{10.1103/PhysRevD.69.046002}{Phys.~Rev.~D69,~046002~(2004)}},
\texttt{\arxivref{hep-th/0305116}{hep-th/0305116}}.

%\%CITATION = HEP-TH/0502226;\%\%
\bibitem{Beisert:2005bm}
N.~Beisert, V.~Kazakov, K.~Sakai and K.~Zarembo,
\textit{``{The Algebraic curve of classical superstrings on AdS(5) x S**5}''},
\textsf{\doiref{10.1007/s00220-006-1529-4}{Commun.Math.Phys.~263,~659~(2006)}},
\texttt{\arxivref{hep-th/0502226}{hep-th/0502226}}.

%\%CITATION = ARXIV:1204.2366;\%\%
\bibitem{deLeeuw:2012jf}
M.~de~Leeuw, T.~Matsumoto, S.~Moriyama, V.~Regelskis and A.~Torrielli,
\textit{``{Secret Symmetries in AdS/CFT}''},
\textsf{\doiref{10.1088/0031-8949/86/02/028502}{Phys.Scripta~02,~028502~(2012)}},
\texttt{\arxivref{1204.2366}{arxiv:1204.2366}}.

\end{thebibliography}

\end{document}